# The Sunyaev-Zel'dovich Effect as a Probe of $\Omega_o$

Domingos Barbosa [1,2], James G. Bartlett [1], Alain Blanchard [1], Jamila Oukbir [3]

[1] *Observatoire Astronomique, 11, rue de l'Université, 67000 Strasbourg, France.*
[2] *Centro de Astrofísica da U.P., Rua do Campo Alegre 823, 4150 Porto, Portugal.*
[3] *SAp, CE-Saclay, 91191 Gif-sur-Yvette Cedex, France.*

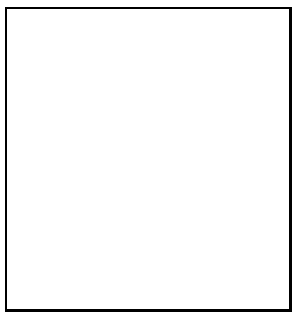

## Abstract

We examine how the Sunyaev-Zel'dovich (SZ) effect of clusters can probe $\Omega_o$. Using self-consistent models of X-ray clusters in the context of hierarchical models of structure formation, we show that both the mean Compton $y$ parameter and the number of clusters observed via the SZ effect strongly depend on $\Omega_o$. These quantities are higher in open cosmological models due to the earlier epoch of structure formation. Specifically, we compare two models which are able to reproduce the present abundance of X-ray clusters, one for $\Omega_o = 0.2$ and the other $\Omega = 1$. For $\Omega_o = 0.2$, $<y>$ exceeds the current FIRAS limit of $1.5 \times 10^{-5}$. Also, the SZ redshift source distribution considerably differs in the two cosmological models. Thus, these results show that both an improvement in our knowledge of CMB spectral distortions and the realization of millimeter surveys to look for SZ sources over a large area of the sky should produce interesting constraints on $\Omega_o$ and on the evolution of the baryonic fraction of virialized structures.

[1] *email: barbosa@wirtz.u-strasbg.fr*

# 1 Introduction

In open cosmological models, structure formation starts earlier, with its rate approaching a constant for redshifts $z_f \sim 1/\Omega_o - 1$. Thus, the structure observed today already existed at $z_f$. An interesting feature is that structure formation rates are independent of the specified spectrum [12]. So, cluster distribution functions should present this caracteristic $\Omega_o$ dependence. Clusters of galaxies are known to have hot intra-cluster gas which emits in X-rays and is the main reservoir of the baryons that fell into the cluster potential wells. This gas will produce a unique spectral distortion in the CMB spectrum by its interaction with CMB photons by inverse Compton scattering [18].

The change, for a given frequency and for a given line of sight, in the sky brightness relative to the CMB intensity, which we will call the SZ surface brightness of a source, is given by $S_\nu = y j_\nu(x)$, where $j_\nu(x)$ describes the frequency dependence and the Compton $y$ parameter measures the distortion magnitude. It has been shown that the integrated SZ surface brightness of a source (the total increment or decrement in the normal CMB flux) depends linearly on the total hot gas content of the cluster and therefore only on their baryonic fraction [2] and not on the spatial distribution of the cluster hot gas :

$$S_\nu = (8 \text{ mJy } h^{8/3}) f_\nu(x) f_B \Omega_o^{1/3} M_{15}^{5/3} [\frac{\Delta(z)}{178}]^{1/3} (1+z) D^{-2}(z), \qquad (1)$$

where $f_B$, z, and $M_{15}$ are the baryonic fraction, redshift and mass of the cluster. $D_a$ and $f_\nu(x)$ are respectively the dimensionless parts of the angular distance and of the characteristic SZ frequency function. We have assumed h=0.5 in all subsequent calculations.

# 2 X-ray Data and the Compton $y$ parameter

In hierarchical models like the cold dark matter (CDM) family of models of structure formation, it is natural to adopt the Press and Schechter (PS) mass function [17] :

$$\frac{dn}{d\ln M} d\ln M = \sqrt{\frac{2}{\pi}} \frac{\rho}{M} \nu(M,z) \left(-\frac{d\ln \sigma}{d\ln M}\right) e^{-\nu^2/2} d\ln M, \qquad (2)$$

where $\rho$ is the the present mass density of the Universe, $\nu = \delta_c D_g(z)/\sigma(M)$, where $\delta_c D_g(z)$ and $\sigma(M)$ are, respectively, the linear density contrast for collapse at an epoch z and the amplitude of the mass fluctuations. The mass variance $\sigma(M)$ is expressed in the Fourier plane by :

$$\sigma(M) = \frac{1}{2\pi^2} \int_0^\infty P(k) W_F^2(kR) k^2 dk. \qquad (3)$$

Here, $W_F$ is some window function and M is the mass contained in a sphere of radius R. In the case of a CDM-like perturbation spectrum, the original primordial power-law spectrum ($P_p$) is modified by the presence of cold, non-baryonic dark matter, among other factors. The change in the spectrum is contained in the CDM transfer function, for which $P(k) = T^2 P_p(k)$, where $P_p \propto k$ just describes the scalar invariant primordial perturbations. A good fitting formula to T(k) is provided by equation(4), where its shape is controled only by the so called $\Gamma$ factor [3] :

$$T(q) = \frac{log(1+2.34q)}{(2.34q)} \left[1 + 3.89q + (16.1q)^2 + (5.46q)^3 + (6.71q)^4\right]^{-1/4}, \qquad (4)$$

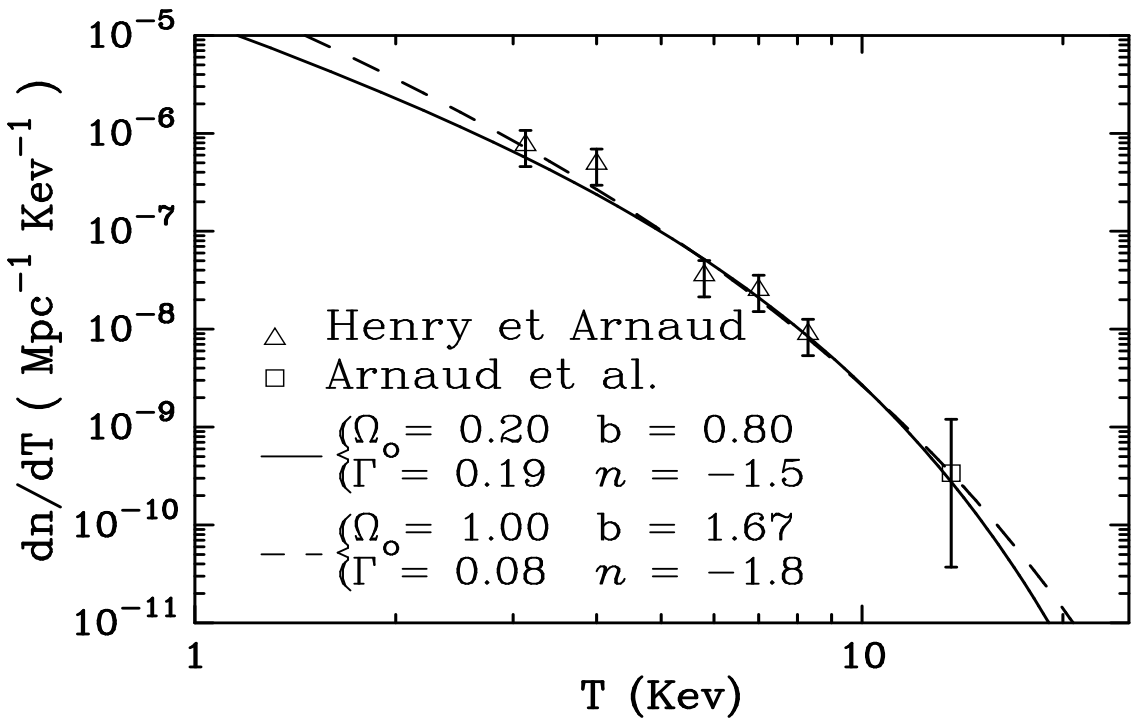

Figure 1: The temperature function for the two models; in both cases we have assumed $h = 0.5$. For each $\Omega_o$, the curves correspond to the best $\chi^2$ fits to $b$ and $\Gamma$. The local power spectrum index is given by $n$.

where $q = k/(h\Gamma)$. In the CDM family, $\Gamma$ is usually identified as $\Omega h$, but here is considered as a free parameter. Also, for the spectrum normalization, we take the bias $b = 1/\sigma(M)$ at $8h^{-1}$ Mpc .

Because clusters are virialized structures, the mass-temperature relation derived in the framework of the spherical collapse model, $T \propto M^{2/3}(1+z)$ [7], together with the mass function yields the temperature function dn/dT. To compare the X-ray data [12], [9], to the theoretical temperature functions for different cosmological models, a $\chi^2$ fit was performed assuming as free parameters the normalization $b$ and the shape of the power spectrum $\Gamma$.

Figure 1 shows the temperature function dn/dT, with the best $\chi^2$ fits to the data for the $\Omega_o = 1$ and $\Omega_o = 0.2$ cosmological models analysed here. An analogue method was used by Oukbir et al. [13], [14] who made a similar $\chi^2$ fit to X-ray data, considering the perturbation spectrum around cluster scales as a power law. They chose as free parameters $b$ and the local power spectral index $n$. The values found by both analyses agree very well, with the $b - \Gamma$ fit indicating the same local spectrum index found by Oukbir et al.

The mean global Compton distortion, $< y >$, will be the sum of all single distortions produced by each cluster. Naturaly, it will be defined as an integral over the entire mass distribution function (in this case the PS mass function) :

$$< y > = f_B \int dz \frac{dt}{dz} c\sigma_T (1+z)^3 \int d\ln M \frac{dn}{d\ln M} \frac{\chi M}{m_p} \frac{kT(M,z)}{m_e c^2}, \qquad (5)$$

where $\chi$ is the number of electrons per baryon. For $f_B$, X-ray observations indicate values as high as $0.05h^{-3/2}$ [6], [8], [19], which agree with nucleosynthesis for $\Omega_o = 0.2 - 0.3$, but are inconsistent for $\Omega_o = 1$ [20]. Also, note that gas in low mass structures is able to cool in less than one Hubble time, so the mass integral in equation(5) has a low mass cut-off. This cut-off is $10^{12} M_\odot$ and is almost independent of redshift [5]. Typically, at redhifts higher than $z \sim 5$, Compton cooling is efficient, so gas within the structures will cool and, consequently, the contribution at these high redshifts to $<y>$ will be small.

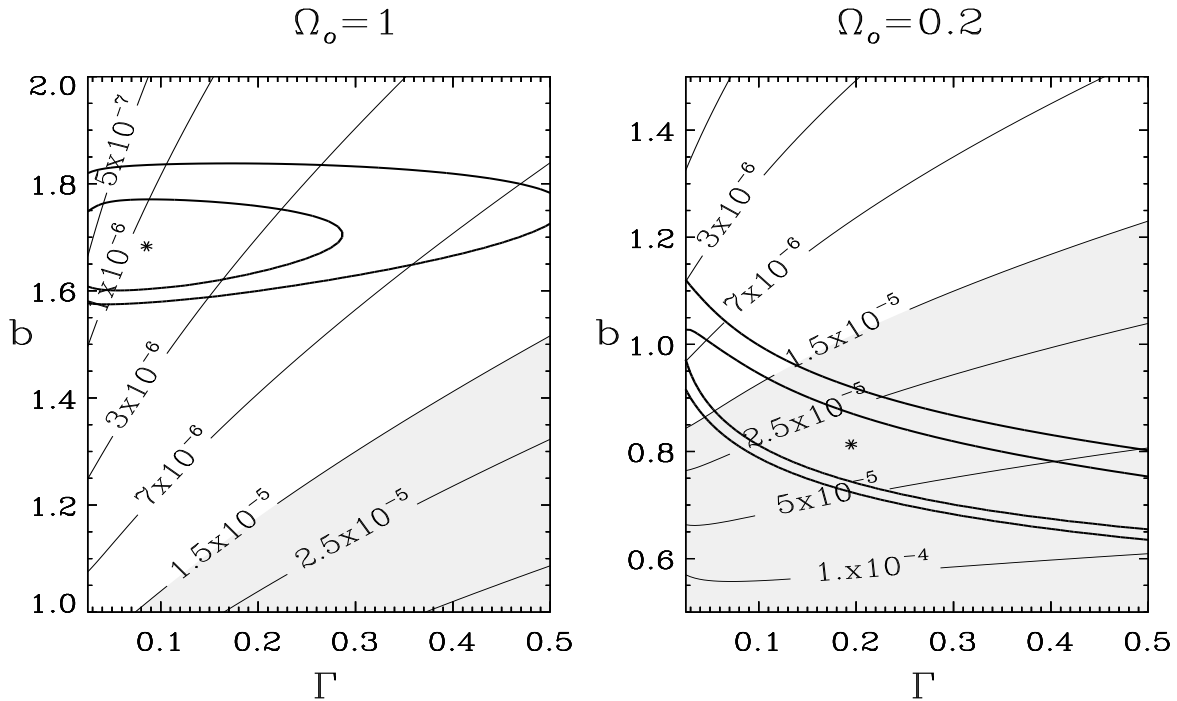

Figure 2: $b - \Gamma$ plane. Thin lines represent $y$ isocontours. The gray zone is the forbidden region : $<y>$ is higher than the current FIRAS limit of $1.5 \times 10^{-5}$. Thick contours represent the $1\sigma$ and the $2\sigma$ contours of the $\chi^2$ fit to temperature function data if errors were gaussian. Asterisks are the best fit values. Note the change in the bias scale between the two panels.

Figure 2 shows X-ray and $<y>$ constraints in the $b - \Gamma$ plane. The ellipses (thicker contours) represent the $1\sigma$ and $2\sigma$ contours of the $\chi^2$ fits. The bias is quite well constrained, unlike $\Gamma$. This is not surprising, because $b$ is defined at the cluster scale, while $\Gamma$ is a shape parameter linking all scales. The best fit values for $\Gamma$ are quite different from the value of $\sim 0.25$ founded by the galaxy distribution [16] but still largely compatible within the $1\sigma$ confidence level. It is clear that for $\Omega_o = 0.2$, only a small region of the $b - \Gamma$ region allowed by X-ray data does not violate the FIRAS limit of $1.5 \times 10^{-5}$ [15]. Further improvement in the limit on $<y>$ (a factor of 2) could provide stronger constraints on this model.

## 3  Source Counts

Integrating the PS mass function, we can study the redshift distribution function of sources brighter on the sky than a certain threshold, $dN/d\Omega/dz(>S_\nu)$, and the the corresponding SZ source counts. The source counts are expressed by :

$$\frac{dN}{d\Omega}(>S_\nu) = \int dz \frac{dV}{dzd\Omega} \int_{M_{min}(S_\nu,z)} dM \frac{dn}{dM}, \qquad (6)$$

where $M_{min}$ is given by equation(1). Figure 3 presents the calculated redshift distribution and SZ source counts at 0.75 mm. The panel on the left shows how different the redshift distribution of the two cosmological models can be : in an open universe, the long high-redshift tail reveals the geometrical effects in the rate of structure formation absent in the flat model. This effect is independent of cluster evolution and it is also visble in X-ray redshift temperature function analysis, [11],[13]. The righthand panel shows the SZ source counts. A difference of 10 is visible at 50 mJy, the expected threshold for COBRAS/SAMBA. If we use a more conservative $\Gamma$ value of 0.25, the difference in the two distributions is still higher than a factor of 4, thus showing the $\Omega_o$ effects.

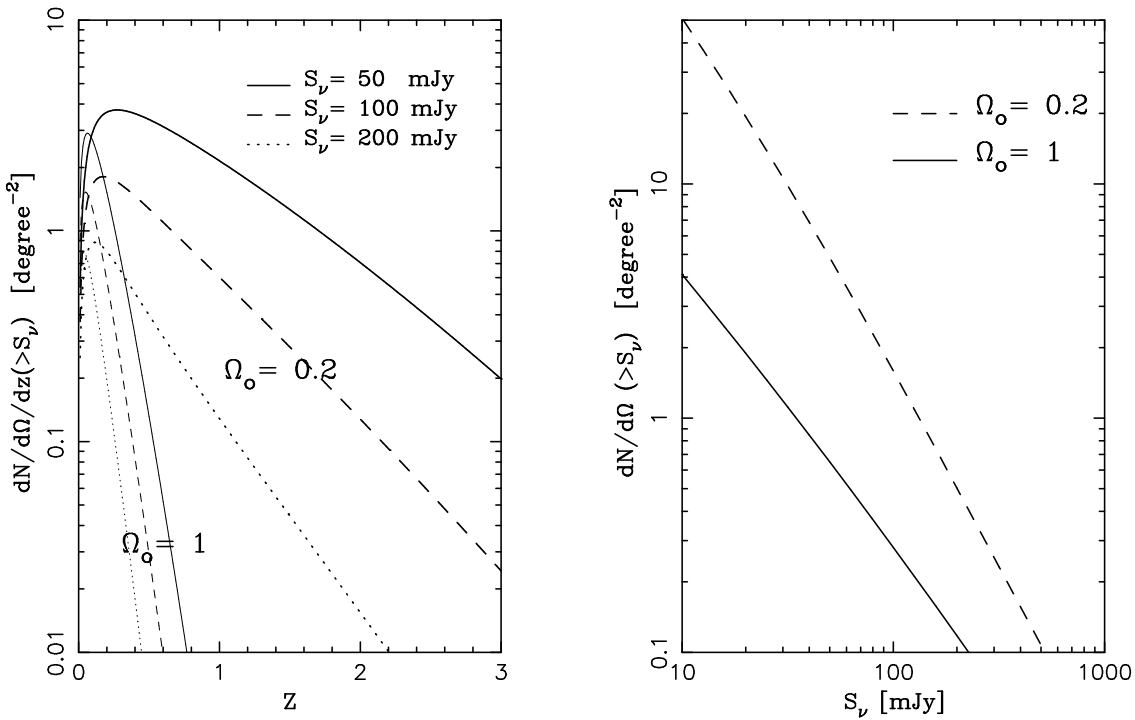

Figure 3: The SZ source counts and the redshift distribution for several $S_\nu$. The parameters $b$ and $\Gamma$ used are the best fit values to dn/dT. The calculation is for a wavelength of 0.75 mm.

## 4 Conclusion

Unlike X-rays, the SZ effect is insensitive to the radial distribution of hot gas in clusters; it depends only on the cluster baryonic fraction. Thus, the Sunyaev-Zel'dovich effect is a good probe of the Universe's geometry. Both the $<y>$ and the SZ source counts show large differences between the flat and open cosmological models considered here. It is clear that an $\Omega = 0.2$ universe which reproduces the temperature cluster function is left with a small region in the $b - \Gamma$ plane. This suggests that an improvement in the limit on $<y>$ should be very constraining for open models. The redshfit distribution of SZ sources is an elegant method of probing $\Omega_o$, due to its independence on the normalisation of the perturbation spectrum. Also, as first pointed out by Korolev et al. [10], cluster source counts should also be a good probe of $\Omega_o$, even if they are more sensitive to evolution than the SZ redshift distribution $dN/d\Omega dz$. In fact, the construction of a catalogue of SZ sources could reveal the $\Omega_o$ geometrical effects as presented in Figure 3.

**Acknowledgements.** D.B is supported by the Praxis XXI CIENCIA-BD/2790/93 grant attributed by JNICT, Portugal.